\documentclass[12pt,showpacs,showkeys,preprintnumbers,amsmath,amssymb,amsfonts,amsthm,amscd,latexsym,prd]{revtex4-1}  %,twocolumn
\usepackage{bm} 
\usepackage{graphicx}
\def\beq{\begin{equation}}
\def\eeq{\end{equation}}
\def\leftcontractp#1{\mathop{\rlap{\hbox to 8pt{\hss$^{#1}$\hss}}%
  \hbox{\vrule height0.5pt width8pt \vrule width0.5pt  height6pt}}}
 
\def\rightcontractp#1{\mathop{%
  \hbox{\vrule width0.5pt height6pt \vrule height0.5pt width8pt}%
    \llap{\hbox to 8pt{\hss$^{#1}$\hss}}}}
 
\input prepictex
\input pictex
\input postpictex

\def\mathput#1{\relax \ifmmode \displaystyle #1\else $\displaystyle #1$\fi}
\def\pmb#1{\setbox0=\hbox{$#1$}%
  \kern-.025em\copy0\kern-\wd0
  \kern.05em\copy0\kern-\wd0
  \kern-.025em\raise.0433em\box0}

\begin{document}

\title{Weitzenb\"ock's  Torsion,  Fermi Coordinates and Adapted Frames}

\author{Donato Bini}
  \affiliation{
Istituto per le Applicazioni del Calcolo ``M. Picone,'' CNR, I-00185 Rome, Italy\\
ICRA, ``Sapienza" University of Rome, I-00185 Rome, Italy
}

\author{Bahram Mashhoon}
  \affiliation{Department of Physics and Astronomy,\\
University of Missouri, Columbia, Missouri 65211, USA
}
\date{\today}

\begin{abstract}
We study Weitzenb\"ock's torsion and discuss its properties. Specifically, we calculate the measured components of  Weitzenb\"ock's torsion tensor for a frame field adapted to static observers in a Fermi normal coordinate system that we establish along the world line of an arbitrary accelerated observer in general relativity.  A similar calculation is carried out in the standard Schwarzschild-like coordinates for static observers in the exterior Kerr spacetime; we then compare our results with the corresponding curvature components. Our work supports the contention that in the extended general relativistic framework involving both the Levi-Civita and Weitzenb\"ock connections, curvature and torsion provide complementary representations of the gravitational field. 
\end{abstract}

\maketitle

\section{Introduction}

It is possible to extend the pseudo-Riemannian (i.e., Lorentzian) structure of General Relativity (GR) in a natural way by adding a second nonsymmetric connection
due to Weitzenb\"ock~\cite{We}. The Weitzenb\"ock connection is related to congruent frames adapted to observer families~\cite{LB}. The standard Levi-Civita connection (${}^0\Gamma^\mu_{\alpha\beta}$) is symmetric and hence torsion-free, but gives rise to the Riemannian curvature of spacetime that characterizes the gravitational field in GR.
On the other hand, the Weitzenb\"ock connection ($\Gamma^\mu_{\alpha\beta}$), which is compatible with the Riemannian metric ($g_{\mu\nu}$), is curvature-free, but has torsion. The curvature of the Levi-Civita connection and the torsion of the Weitzenb\"ock connection are complementary aspects of the gravitational field in the recent nonlocal generalization of GR~\cite{HM1, HM2, RM, Mas}.

In a global inertial frame in Minkowski spacetime, there exists a natural system of globally parallel tetrad frames, since the ideal inertial observers at rest carry orthonormal tetrad frames that consist of the four unit basis vectors of the background inertial frame in which the observers are all at rest. Flat spacetime contains an equivalence class of such \emph{parallel} frame fields that are related to each other by 
constant  elements of the six-parameter global Lorentz group.  This parallelism disappears in the curved spacetime of GR. That is, given any smooth orthonormal tetrad field $\lambda^\mu{}_{\hat \alpha}(x)$ adapted to an observer family in curved spacetime, it is not possible to render the frame field parallel in any spacetime domain due to the presence of the Riemannian curvature of the Levi-Civita connection. It is nevertheless useful to have access to a global system of \emph{parallel} axes in the presence of gravitation. To this end, one must extend GR by introducing a second (Weitzenb\"ock) connection, which is so defined as to render a smooth orthonormal frame field parallel  in extended GR. Therefore, of all possible smooth frame fields on Riemannian spacetime, one system can be chosen in order to define a global system of parallel axes that are, however, specified up to global Lorentz transformations. This circumstance is reminiscent of the parallel frame fields of \emph{inertial} observers in Minkowski spacetime. In extended GR, the chosen parallel frame field is adapted to a \emph{preferred} family of observers. Such an observer family is then unique up to global Lorentz transformations, just as is the case with inertial observers in Minkowski spacetime. Henceforth, a \emph{preferred} observer family in extended GR is one for which the frame field is globally \emph{parallel} via the Weitzenb\"ock connection. 

Imagine a class of preferred observers in extended GR and their associated smooth tetrad frame field $e^\mu{}_{\hat \alpha}$ such that
\beq
\label{eq:1}
g_{\mu\nu}\,e^\mu{}_{\hat \alpha}\,e^\nu{}_{\hat \beta}=\eta_{\hat \alpha \hat \beta}\,,
\eeq
which is the orthonormality condition for the frame field. For this class of preferred observers, the Weitzenb\"ock connection is given by~\cite{We}
\beq
\label{eq:2}
\Gamma^\mu_{\alpha\beta}=e^\mu{}_{\hat \nu}\,\partial_\alpha\, e_\beta{}^{\hat \nu}\,.
\eeq
This is essentially the unique connection for which the corresponding covariant differentiation is such that $\nabla_\nu\, e_\mu{}^{\hat \alpha}=0$. From this and the orthonormality condition, we get metric compatibility; that is, $\nabla_\nu\, g_{\alpha\beta}=0$. This leads to a global notion of parallelism; namely, distant vectors may be considered parallel if they have the same local components relative to their preferred frames. \emph{Teleparallelism} has a long history~\cite{HS, BH, AP, Mal}; in this framework, GR has an equivalent teleparallel formulation (GR$_{||}$)~\cite{Mo, PP}.

The difference between two connections on the same manifold is a tensor. Thus we have the \emph{torsion} tensor
\beq
\label{eq:3}
C_{\alpha \beta}{}^\mu = \Gamma^\mu_{\alpha\beta}-\Gamma^\mu_{\beta\alpha}=e^\mu{}_{\hat \nu}\left(
\partial_\alpha\, e_\beta{}^{\hat \nu}-\partial_\beta\, e_\alpha{}^{\hat \nu}\right)\,,
\eeq
and the \emph{contorsion} tensor
\beq
\label{eq:4}
K_{\alpha \beta}{}^\mu = {}^0\Gamma^\mu_{\alpha\beta}-\Gamma^\mu_{\alpha\beta}\,.
\eeq
It follows from the compatibility of the Levi-Civita and Weitzenb\"ock connections with the Riemannian metric that  the contorsion tensor is linearly related to the torsion tensor via
\begin{equation}\label{eq:5}
K_{\alpha \beta \gamma} = \frac{1}{2} (C_{\alpha \gamma \beta}+C_{\beta \gamma \alpha}-C_{\alpha \beta \gamma})\,.
\end{equation}
We note that the torsion tensor is antisymmetric in its first two indices, while the contorsion tensor is antisymmetric in its last two indices. In this paper, we choose units such that $G=c=1$. Furthermore, Greek indices run from 0 to 3, while Latin indices run from 1 to 3. The signature of the metric is +2. We use a left superscript  ``0" for geometric quantities related to the Levi-Civita connection. Our conventions regarding the use of a nonsymmetric connection are explained in Appendix A. 

Let us consider the frame components of the Weitzenb\"ock torsion with respect to the preferred  orthonormal frame $e_{\hat \alpha}$ with dual $\omega^{\hat \alpha}$ such that $\omega^{\hat \alpha}(e_{\hat \beta})=\delta^{\hat\alpha}_{\hat \beta}$; that is,
\beq \label{eq:6}
C_{\hat \alpha \hat \beta}{}^{\hat \gamma}=e^\mu{}_{\hat \alpha}\,e^\nu{}_{\hat \beta}\left(\partial_\mu\, e_\nu{}^{\hat \gamma}-\partial_\nu\, e_\mu{}^{\hat \gamma}  \right)\,.
\eeq
These are measurable in principle and are essentially the structure functions of the preferred frame $e_{\hat \alpha}=e^\mu{}_{\hat \alpha}\,\partial_\mu$; that is,   
\beq \label{eq:7}
[e_{\hat \alpha}\,, e_{\hat \beta}] = - C_{\hat \alpha \hat \beta}{}^{\hat \gamma}\,e_{\hat \gamma}\,.
\eeq
Equivalently, these components can be obtained by evaluating the exterior derivative of the frame $1$-forms $\omega^{\hat \gamma}=e_\mu{}^{\hat \gamma}\, dx^\mu$ according to the relation
\beq \label{eq:8}
d \omega^{\hat \gamma}=\frac12 C_{\hat \alpha \hat \beta}{}^{\hat \gamma}\, \omega^{\hat \alpha}\wedge \omega^{\hat \beta}\,,
\eeq
or the Lie derivative of the frame vectors along each other
\beq  \label{eq:9}
\pounds_{e_{\hat \alpha}}\, e_{\hat \beta}=[e_{\hat \alpha}\,, e_{\hat \beta}]=-C_{ \hat \alpha \hat \beta }{}^{\hat \gamma}\,e_{\hat \gamma}\,
\eeq
and its  ``dual" relation
\beq  \label{eq:10}
\pounds_{e_{\hat \alpha}}\, \omega^{\hat \beta}=C_{ \hat \alpha \hat \gamma }{}^{\hat \beta} \omega^{\hat \gamma}\,.
\eeq

At any event in spacetime, two orthonormal frames are related to each other by an element  of the local Lorentz group; therefore, $C_{ \hat \alpha \hat \beta }{}^{\hat \gamma}$ transforms as a third-rank tensor under local Lorentz transformations. Moreover, these structure functions satisfy the Jacobi identity, 
\beq \label{eq:11}
[e_{\hat \alpha}\,, [e_{\hat \beta}\,, e_{\hat \gamma}]\,] + [e_{\hat \beta}\,, [e_{\hat \gamma}\,, e_{\hat \alpha}]\,] + [e_{\hat \gamma}\,, [e_{\hat \alpha}\,, e_{\hat \beta}]\,]=0\,,
\eeq
which is equivalent to $d^2\omega^{\hat \alpha}=0$. It follows from the Jacobi identity that
\begin{eqnarray}  \label{eq:12}
\partial_{\,[\,\hat \alpha}\, C_{\hat \beta \hat \gamma]\,}{}^{\hat \mu}+C_{\hat \sigma [\hat \alpha}{}^{\hat \mu} \,C_{\hat \beta \hat \gamma]}{}^{\hat \sigma}=0\,,
\end{eqnarray}
where $\partial_{\hat \alpha}:=e_{\hat \alpha}$ is the Pfaffian derivative associated with $e_{\hat \alpha}$.

The main purpose of this paper is to calculate the structure functions $C_{ \hat \alpha \hat \beta }{}^{\hat \gamma}$ in a general and physically transparent setting and study their physical properties. The following section is devoted to the study of the structure functions in the physically meaningful Fermi coordinates in a general gravitational field. In section III, we examine the structure functions for static observers in a general stationary axisymmetric gravitational field such as the Kerr spacetime. Section IV is devoted to a brief discussion of the lack of closure of \emph{infinitesimal} parallelograms in the presence of torsion. Finally, section V contains a discussion of our results. 
\section{Weitzenb\"ock's Torsion in Fermi Coordinates}

To gain physical insight into the structure of Weitzenb\"ock's torsion, we consider an arbitrary gravitational field in extended GR and establish a Fermi coordinate system in a cylindrical spacetime region along the world line of an arbitrary accelerated observer ${\cal O}$. Fermi coordinates are invariantly defined and constitute the natural general-relativistic generalization of inertial Cartesian coordinates. We then define the frame field of \emph{static} observers in the Fermi coordinate system and calculate explicitly their measured torsion tensor $C_{\hat \alpha \hat \beta}{}^{\hat \gamma}$.

Imagine an accelerated observer ${\cal O}$ following the reference world line ${\bar x}^\mu (\tau)$, where $x^\mu= (t, x^i)$ is an admissible system of spacetime coordinates~\cite{BCM} and $\tau$ is the proper time along the observer's trajectory. The observer carries  an orthonormal tetrad frame $\lambda^{\mu}{}_{\hat{\alpha}}(\tau )$ along its path in accordance with
\begin{equation}\label{F1} 
\frac{^0D\lambda^\mu{}_{\hat{\alpha}}}{d\tau} =\phi_{\hat{\alpha}}{}^{\hat{\beta}}(\tau)~ 
\lambda^\mu{}_{\hat{\beta}}\,.
\end{equation}
Here, $\phi_{\hat{\alpha} \hat{\beta}}=-\phi_{\hat{\beta} \hat{\alpha}}$ is the acceleration tensor of ${\cal O}$. In close analogy with the Faraday tensor, we can decompose the acceleration tensor into its ``electric" and ``magnetic" components, namely,  $\phi_{\hat{\alpha} \hat{\beta}} \mapsto (-\mathbf{a}, \boldsymbol{\Omega})$. That is, the translational acceleration vector $\mathbf{a}$ is given by the frame components of the 4-acceleration vector associated with the 4-velocity vector  $\lambda^\mu{}_{\hat{0}}=d{\bar x}^\mu/d\tau$ of the observer and $\boldsymbol{\Omega}$ is the angular velocity of the rotation of the observer's local spatial triad $\lambda^{\mu}{}_{\hat{i}}$\,,\, $i=1,2,3$, with respect to the locally nonrotating (i.e., Fermi-Walker transported) triad~\cite{Mash}.

Let us next establish  an extended Fermi normal coordinate system in a world tube along ${\bar x}^\mu (\tau)$. The Fermi coordinates are scalar invariants by construction and are indispensable for the interpretation of measurements in GR---see~\cite{Sy, BMash, NZ, CM1, CM2, CM3} and the references cited therein. 
Consider the class of spacelike geodesics that are orthogonal to the world line of the accelerated observer at each event $Q(\tau)$ along ${\bar x}^\mu (\tau)$. These form a local hypersurface.  
For an event $P$ with coordinates $x^\mu$ on this hypersurface, let there be a \emph{unique} spacelike geodesic of proper length $\sigma$ that connects $Q$ to $P$. Then, $P$ has Fermi coordinates  $X^{\hat \mu}=(T, X^{\hat i})$, where
\begin{equation}\label{F2}
T=\tau\,, \qquad X^{\hat i}=\sigma\, \xi^\mu\, \lambda_{\mu}{}^{\hat{i}}(\tau)\,.
\end{equation}
Here, $\xi^\mu$ is the \emph{unit} vector at $Q(\tau)$ that is tangent to the spacelike geodesic segment from $Q$ to $P$. Thus the reference observer $\cal{O}$ is always at the spatial origin of the Fermi coordinate system.

The coordinate transformation $x^\mu \mapsto (X^{\hat 0}=T,X^{\hat a})$ can only be specified implicitly in general; hence, it is useful to express the spacetime metric in Fermi coordinates as a Taylor expansion in powers of the spatial distance $\sigma$ away from the reference world line. For our present purposes, we can write the metric in Fermi coordinates as
\begin{eqnarray}
\label{F3}
g_{\hat 0 \hat 0}&=& -{\cal P}^2+{\cal Q}^2 -  R_{\hat 0 \hat i\hat 0 \hat j}X^{\hat i} X^{\hat j}+O(|\mathbf{X}|^3)\,,\nonumber\\
&=& -{\cal P}^2+{\cal Q}^2 -2\Phi+O(|\mathbf{X}|^3)\,,\nonumber\\
g_{\hat 0 \hat i}&=& {\cal Q}_{\hat i} -\frac23   R_{\hat 0 \hat j \hat i \hat k}X^{\hat j}X^{\hat k} +O(|\mathbf{X}|^3)\,,\nonumber\\
&=&  {\cal Q}_{\hat i}-2{\mathcal A}_{\hat i} +O(|\mathbf{X}|^3)\,,\nonumber\\
g_{\hat i \hat j}&=& \delta_{\hat i \hat j}-\frac13   R_{\hat i \hat k \hat j \hat l}X^{\hat k}X^{\hat l}+O(|\mathbf{X}|^3)\,,\nonumber\\
&=& \delta_{\hat i  \hat j}-2\Sigma_{\hat i  \hat j}+O(|\mathbf{X}|^3)\,.
\end{eqnarray}
Here, we have introduced
\beq \label{F4}
{\cal P}= 1+ U\,,\qquad  U={\mathbf a} \cdot {\mathbf X}\,, \qquad \mathbf{{\cal Q}}= \boldsymbol{\Omega} \times {\mathbf X} \,
\eeq
and we have used the notation
\beq \label{F5}
\Phi=\frac12  R_{\hat 0 \hat i \hat 0 \hat j}X^{\hat i} X^{\hat j}\,,\qquad
{\mathcal A}_{\hat i} =\frac13   R_{\hat 0 \hat j \hat i \hat k}X^{\hat j}X^{\hat k} \,,
\qquad \Sigma_{\hat i \hat j}=\frac16 R_{\hat i \hat k \hat j \hat l}X^{\hat k}X^{\hat l}\,.
\eeq
Moreover,  $R_{\hat \alpha \hat \beta \hat \gamma \hat \delta}(T)$ is the projection of the 
Riemann curvature tensor on the orthonormal tetrad frame of ${\cal O}$ and evaluated along the reference geodesic; that is,
\begin{equation}\label{F6} 
R_{\hat \alpha \hat \beta \hat \gamma \hat \delta}(T):=\,^{0}R_{\mu\nu \rho
\sigma}\,\lambda^\mu{}_{\hat{\alpha}}\,\lambda^\nu{}_{\hat{\beta}}\,\lambda^\rho{}_{\hat{\gamma}}\,\lambda^\sigma{}_{\hat{\delta}}\,.
\end{equation}
Henceforward, we will only keep terms up to second-order in the metric perturbation and note that Fermi coordinates are admissible in a finite cylindrical region about the world line of ${\cal O}$ with $|\mathbf{X}|\ll r_c$, where $r_c(T)$ is the infimum of acceleration lengths $(|\mathbf{a}(T)|^{-1}, |\boldsymbol{\Omega}(T)|^{-1})$ as well as spacetime curvature lengths such as $|R_{\hat \alpha \hat \beta \hat \gamma \hat \delta}(T)|^{-1/2}$. 

Let us now consider the class of observers that are all at \emph{rest} in this gravitational field and carry orthonormal tetrads that have essentially the same orientation as the Fermi coordinate system. This class includes of course our reference observer ${\cal O}$. The orthonormal tetrad frame of these preferred observers can be expressed in $(T, X^{\hat i})$ coordinates as 
\begin{align}
\label{F9} e^{\mu}{}_{\hat{0}}&= (1-\tilde{\Phi},\, 0,\, 0,\, 0)\,,\\
\label{F10} e^{\mu}{}_{\hat{1}}&=(-2\tilde{{\cal A}}_{\hat 1},\, 1+\tilde{\Sigma}_{\hat 1 \hat 1},\, 0,\, 0)\,,\\
\label{F11} e^{\mu}{}_{\hat{2}}&=(-2\tilde{{\cal A}}_{\hat 2},\, 2\,\tilde{\Sigma}_{\hat 2 \hat 1},\, 1+\tilde{\Sigma}_{\hat 2 \hat 2},\, 0)\,,\\
\label{F12} e^{\mu}{}_{\hat{3}}&=(-2\tilde{{\cal A}}_{\hat 3},\, 2\,\tilde{\Sigma}_{\hat 3 \hat 1},\, 2\,\tilde{\Sigma}_{\hat 3 \hat 2},\, 1+\tilde{\Sigma}_{\hat 3 \hat 3})\,.
\end{align}
Here, we have defined
\beq \label{F13}
\tilde{\Phi} := \Phi + U-U^2 -\frac{1}{2}{\cal Q}^2\,,\quad \tilde{{\cal A}}_{\hat i}:={\cal A}_{\hat i}-(\frac{1}{2}-U){\cal Q}_{\hat i}\,, \quad \tilde{\Sigma}_{\hat i \hat j}:=\Sigma_{\hat i \hat j}-\frac{1}{2} {\cal Q}_{\hat i}{\cal Q}_{\hat j}\,.
\eeq
As expected, $e^{\mu}{}_{\hat{\alpha}}$ reduces to $\delta^\mu_{\hat \alpha}$ along the reference geodesic, where $\mathbf{X}=0$. It follows from $e_{\mu \hat{\alpha}}=g_{\mu \nu} \,e^{\nu}{}_{\hat{\alpha}}$ that 
\begin{align}
\label{F16} e_{\mu\, \hat{0}}&= (-1-U^2-\tilde{\Phi},\, -2\tilde{{\cal A}}_{\hat 1}+U{\cal Q}_{\hat 1},\, -2\tilde{{\cal A}}_{\hat 2}+U{\cal Q}_{\hat 2},\, -2\tilde{{\cal A}}_{\hat 3}+U{\cal Q}_{\hat 3})\,,\\
\label{F17} e_{\mu\, \hat{1}}&=(0,\, 1-\tilde{\Sigma}_{\hat 1 \hat 1},\, -2\,\tilde{\Sigma}_{\hat 1 \hat 2},\, -2\,\tilde{\Sigma}_{\hat 1 \hat 3})\,,\\
\label{F18} e_{\mu\, \hat{2}}&=(0,\, 0,\, 1-\tilde{\Sigma}_{\hat 2 \hat 2},\, -2\,\tilde{\Sigma}_{\hat 2 \hat 3})\,,\\
\label{F19} e_{\mu\, \hat{3}}&=(0,\, 0,\, 0,\, 1-\tilde{\Sigma}_{\hat 3 \hat 3})\,.
\end{align}
Explicitly, we therefore have
\begin{eqnarray} \label{F20}
e_{\hat 0}&=& \left(1-U+U^2+\frac12 {\cal Q}^2-\Phi  \right) \partial_T\nonumber\,,\\
e_{\hat 1}&=& \left[ -2{\mathcal A}_{\hat 1}+{\cal Q}_{\hat 1}(1-2U) \right]  \partial_T 
+ \left(1-\frac12 {\cal Q}_{\hat 1}^2+\Sigma_{\hat 1 \hat 1}  \right)  \partial_{X^{\hat 1}}\,,\nonumber\\
e_{\hat 2}&=&  \left[ -2{\mathcal A}_{\hat 2}+{\cal Q}_{\hat 2}(1-2U)   \right] \partial_T
+ \left(2\Sigma_{\hat 2 \hat 1}-{\cal Q}_{\hat 2}{\cal Q}_{\hat 1}  \right)  \partial_{X^{\hat 1}} + \left( 1-\frac12 {\cal Q}_{\hat 2}^2+\Sigma_{\hat 2 \hat 2}  \right) \partial_{X^{\hat 2}}\,,\nonumber\\
e_{\hat 3}&=& \left[-2{\mathcal A}_{\hat 3}+{\cal Q}_{\hat 3}(1-2U)  \right]   \partial_T
+ \left(2\Sigma_{\hat 3 \hat 1}-{\cal Q}_{\hat 3}{\cal Q}_{\hat 1}    \right)  \partial_{X^{\hat 1}}+ \left( 2\Sigma_{\hat 3 \hat 2}-{\cal Q}_{\hat 3}{\cal Q}_{\hat 2}   \right)  \partial_{X^{\hat 2}}\nonumber\\
&& 
+ \left( 1-\frac12 {\cal Q}_{\hat 3}^2+\Sigma_{\hat 3 \hat 3}  \right)  \partial_{X^{\hat 3}}
\end{eqnarray}
with dual frame
\begin{eqnarray} \label{F21}
\omega^{\hat 0}&=& \left(1+U-\frac12 Q^2+\Phi  \right) dT+[ 2{\mathcal A}_{\hat a}-{\cal Q}_{\hat a}(1-U)]dX^{\hat a}\,,\nonumber\\
\omega^{\hat 1}&=&  
 \left(1+\frac12 {\cal Q}_{\hat 1}^2-\Sigma_{\hat 1 \hat 1}  \right) dX^{\hat 1}
+ \left({\cal Q}_{\hat 1}{\cal Q}_{\hat 2}-2\Sigma_{\hat 1 \hat 2}  \right) dX^{\hat 2}
+ \left({\cal Q}_{\hat 1}{\cal Q}_{\hat 3}-2\Sigma_{\hat 1 \hat 3}  \right) dX^{\hat 3}\,,
\nonumber\\
\omega^{\hat 2}&=&  
 \left(1+\frac12 {\cal Q}_{\hat 2}^2-\Sigma_{\hat 2 \hat 2} \right) dX^{\hat 2}
+ \left({\cal Q}_{\hat 2}{\cal Q}_{\hat 3}-2\Sigma_{\hat 2 \hat 3}  \right) dX^{\hat 3}\,,
\nonumber\\
\omega^{\hat 3}&=&
\left(1+\frac12 {\cal Q}_{\hat 3}^2-\Sigma_{\hat 3 \hat 3}   \right) dX^{\hat 3}\,.
\end{eqnarray}
We can now proceed to the evaluation of the associated structure functions.

In $C_{\hat \alpha \hat \beta}{}^{\hat \gamma}$, for each $\hat \gamma=\hat 0, \hat 1, \hat 2, \hat 3$, we have an antisymmetric tensor that has ``electric" and ``magnetic" components in analogy with the Faraday tensor. Indeed, for $\hat \gamma=\hat 0$, we have 
\begin{equation}\label{F22}
 C_{\hat 0 \hat i}{}^{\hat 0}=-\left[\boldsymbol{{\cal E}}+(1-U)\mathbf{a}+\boldsymbol{\Omega} \times {\boldsymbol{{\cal Q}}}+\dot {\boldsymbol{{\cal Q}}}\right]_{\hat i}\,,  \quad  C_{\hat i \hat j}{}^{\hat 0}=2 \epsilon_{\hat i \hat j \hat k}\left[\boldsymbol{{\cal B}}- (1-U)\,\boldsymbol{\Omega}+{\mathbf a} \times \boldsymbol{{\cal Q}}\right]^{\hat k}\,,
 \end{equation}
 where
 \begin{equation}\label{F22a} 
\dot {\boldsymbol{{\cal Q}}}=\left(\frac{d\boldsymbol{\Omega}}{dT}\right)\times \mathbf{X}\,, \qquad {\mathbf a} \times \boldsymbol{{\cal Q}}=U \boldsymbol{\Omega}-(\mathbf{a}\cdot \boldsymbol{\Omega}) \mathbf{X}\,.
\end{equation}
Furthermore, the gravitoelectric field, $\boldsymbol{{\cal E}}=\boldsymbol{\nabla} \Phi$, and the gravitomagnetic field, $\boldsymbol{{\cal B}}=\boldsymbol{\nabla} \times \boldsymbol{{\cal A}}$\,, are  given by
\begin{equation}\label{F22b} 
{\cal E}_{\hat i}(T ,{\bf X})=R_{\hat 0 \hat i \hat 0 \hat j}(T) X^{\hat j}\,,\qquad
{\cal B}_{\hat i}(T ,{\bf X})=-\frac{1}{2}\epsilon_{\hat i \hat j \hat k}R^{\hat j \hat k}{}_{\hat 0 \hat l}(T )X^{\hat l}\,.
\end{equation}
Let us note here that the gravitoelectric field is directly proportional to the ``electric" components of the Riemann curvature tensor  and similarly the gravitomagnetic 
field is directly proportional to the ``magnetic''
components of the Riemann curvature tensor. It is interesting that we can couch our torsion results in the familiar language of gravitoelectromagnetism (GEM)~\cite{Matt, JCB, Mashhoon}. Moreover, the spatial part of the metric perturbation away from Minkowski spacetime, $\Sigma_{ij}=\Sigma_{ji}$, is likewise proportional to the spatial components of the curvature. Next, for $\hat \gamma=\hat 1, \hat 2, \hat 3$, the electric parts only involve terms of higher order and can be ignored, so that
\begin{equation}\label{F23}
 C_{\hat 0 \hat i}{}^{\hat j}=0\,.
 \end{equation}
However, the corresponding magnetic parts depend upon the spatial components of the curvature and we find that for $\hat \gamma=\hat 1$, 
\begin{equation}\label{F24}
 C_{\hat 2 \hat 3}{}^{\hat 1}=3\Omega_{\hat 1}{\cal Q}_{\hat 1}+ R_{\hat 2 \hat 3 \hat 1  \hat a}X^{\hat a}\,, \quad  C_{\hat 3 \hat 1}{}^{\hat 1}=2 \Omega_{\hat 2}{\cal Q}_{\hat 1}+\frac23 R_{\hat 3  \hat 1 \hat 1 \hat a}X^{\hat a}\,, \quad C_{\hat 1 \hat 2}{}^{\hat 1}=2\Omega_{\hat 3}{\cal Q}_{\hat 1}+\frac23 R_{\hat 1  \hat 2 \hat 1 \hat a}X^{\hat a} \,.
 \end{equation}
Similarly, for $\hat \gamma=\hat 2$,
\begin{eqnarray}\label{F25}
C_{\hat 2 \hat 3}{}^{\hat 2}&=&2\Omega_{\hat 1}{\cal Q}_{\hat 2}+ \frac23 R_{\hat 2 \hat 3 \hat 2  \hat a}X^{\hat a}\,, \quad C_{\hat 3 \hat 1}{}^{\hat 2}=\Omega_{\hat 2}{\cal Q}_{\hat 2} -\Omega_{\hat 3}{\cal Q}_{\hat 3}+\frac13 (R_{\hat 3  \hat 1 \hat 2 \hat a}
- R_{\hat 1  \hat 2 \hat 3 \hat a})X^{\hat a}\,, \nonumber\\
\quad C_{\hat 1 \hat 2}{}^{\hat 2}&=& \Omega_{\hat 3}{\cal Q}_{\hat 2}+\frac13 R_{\hat 1  \hat 2 \hat 2 \hat a}X^{\hat a} \,,
 \end{eqnarray}
and for $\hat \gamma=\hat 3$, 
\begin{equation}\label{F26}
 C_{\hat 2 \hat 3}{}^{\hat 3}= \Omega_{\hat 1}{\cal Q}_{\hat 3}+\frac13  R_{\hat 2 \hat 3 \hat 3  \hat a}X^{\hat a}\,, \quad  C_{\hat 3 \hat 1}{}^{\hat 3}=
\Omega_{\hat 2}{\cal Q}_{\hat 3}+\frac13 R_{\hat 3  \hat 1 \hat 3 \hat a}X^{\hat a} \,, \quad C_{\hat 1 \hat 2}{}^{\hat 3}=0\,.
 \end{equation}
It is important to note that all of the components of $C_{\hat \alpha \hat \beta}{}^{\hat \gamma}$ can be obtained from Eqs.~\eqref{F22}--\eqref{F26} by using the antisymmetry of  $C_{\hat \alpha \hat \beta}{}^{\hat \gamma}$  in its first two indices. Furthermore,  \emph{all} of the components of the curvature tensor are involved in our calculation of the torsion tensor.  The spatial components of the curvature tensor in Eqs.~\eqref{F24}--\eqref{F26} essentially reduce to the gravitoelectric components in a Ricci-flat region of spacetime.  The work reported here generalizes and extends the results of a previous investigation regarding the possibility of  measurement of Weitzenb\"ock's torsion~\cite{BaMa}.

The torsion vector $C_{\alpha}$, $C_\alpha:=-C_{\alpha \beta}{}^\beta$, can be calculated for the static Fermi observers and turns out to be completely spatial; that is, $C_{\hat \alpha}= (0, \Theta_{\hat i})$, where $\boldsymbol{\Theta}$ is related to the gravitoelectric field as well as the spatial part of the torsion tensor. Indeed, 
\begin{equation}\label{F26a} 
\Theta_{\hat i} =-\left[\boldsymbol{{\cal E}}+(1-U)\mathbf{a}+\boldsymbol{\Omega} \times {\boldsymbol{{\cal Q}}}+\dot {\boldsymbol{{\cal Q}}}\right]_{\hat i}-C_{\hat i \hat j}{}^{\hat j}\,.
\end{equation}
On the other hand, the torsion pseudovector  $\check{C}_{\alpha}$, $\check{C}_{\alpha}:=-(1/6) \epsilon_{\hat \alpha \hat \beta \hat \gamma \hat \delta}\,C^{\hat \beta \hat \gamma \hat \delta}$ is given by $(\check{C}_{\hat 0}, H_{\hat i})$, where $\check{C}_{\hat 0}=-(1/3)(C_{\hat 2 \hat 3}{}^{\hat 1}+C_{\hat 3 \hat 1}{}^{\hat 2})$ and $\mathbf{H}$ is related to the gravitomagnetic field,
\begin{equation}\label{F26b} 
\mathbf{H} =\frac{2}{3}\,\left[\boldsymbol{{\cal B}}- (1-U)\,\boldsymbol{\Omega}+{\mathbf a} \times \boldsymbol{{\cal Q}}\right]\,.
\end{equation}
We note that in our convention $\epsilon_{0123}=1$. The three algebraic Weitzenb\"ock invariants of the torsion tensor are discussed in Appendix C. 

It is interesting to compute the acceleration tensor for our family of static observers. To this end, we have 
\begin{equation}\label{F27} 
\frac{^0D\,e^\mu{}_{\hat{\alpha}}}{ds} =\Upsilon_{\hat{\alpha}}{}^{\hat{\beta}}~ 
e^\mu{}_{\hat{\beta}}\,,
\end{equation}
where $s$ is the proper time along the observer's  world line. From Eq.~\eqref{eq:4} and the fact that $e^\mu{}_{\hat 0}=dx^\mu/ds$, we find that
\begin{equation}\label{F28} 
\Upsilon_{\hat{\alpha} \hat{\beta}} =K_{\hat{0} \hat{\alpha} \hat{\beta}}\,.
\end{equation}
Let us briefly digress here and mention that the connection between Eqs.~\eqref{F27} and~\eqref{F28} is completely general and is independent of the particular coordinate system or our choice of the preferred observers. In the particular case of Fermi coordinates and static observers under consideration here, however, it follows from the decomposition of $\Upsilon_{\hat \alpha \hat \beta}$  into its  ``electric" and ``magnetic" components and Eq.~\eqref{eq:5} that $-C_{\hat 0 \hat i}{}^{\hat 0}$ and $-\frac{1}{2}\,C_{\hat i \hat j}{}^{\hat 0}$ are responsible for the proper acceleration and rotation of our observer family, respectively. This circumstance accounts for the nature of the terms that appear in  Eq.~\eqref{F22}, such as, for instance, the centripetal and transverse (Euler) acceleration terms in the electric components. 

The torsion tensor vanishes along the reference world line $(\mathbf {X}=0)$ if $\phi_{\hat \alpha \hat \beta}=0$. It follows that along the reference \emph{geodesic}, the contorsion tensor and the Weitzenb\"ock connection both vanish. Thus, by a proper choice of coordinates and preferred frame field, the Levi-Civita connection as well as the Weitzenb\"ock connection can be made to vanish along a timelike geodesic. This provides a natural generalization of Fermi's result in the context of extended GR.
  
\section{Kerr Spacetime}

Imagine accelerated observers at rest far away from a rotating gravitational source. Within the framework of linearized GR, the spacetime metric in gravitoelectromagnetic (GEM) form is given by~\cite{Mashhoon} 
\begin{equation}\label{K1}
ds^2=-(1+2\Phi')dt^2-4(\mathbf{A}\cdot d\mathbf{x})dt + (1-2\Phi')\delta_{i j}dx^idx^j\,,
\end{equation} 
where $\Phi'$ is the gravitoelectric potential of the source and $\mathbf{A}$ is the corresponding  gravitomagnetic vector potential.  Following the linear perturbation approach, the GEM fields are given by~\cite{Mashhoon}
\begin{equation}\label{K1a}
\mathbf{E}=\boldsymbol{\nabla} \Phi' -\frac{\partial}{\partial t}\left(\frac{1}{2} \mathbf{A}\right)\,, \qquad \mathbf{B}=\boldsymbol{\nabla} \times \mathbf{A}\,.
\end{equation} 
The natural orthonormal tetrad frame adapted to these \emph{static} observers can be expressed as
\begin{align}
\label{K2} e^{\mu}{}_{\hat{0}}&= (1-\Phi',\, 0,\, 0,\, 0)\,,\\
\label{K3} e^{\mu}{}_{\hat{1}}&=(-2A_{\hat 1},\, 1+\Phi',\, 0,\, 0)\,,\\
\label{K4} e^{\mu}{}_{\hat{2}}&=(-2A_{\hat 2},\, 0,\, 1+\Phi',\, 0)\,,\\
\label{K5} e^{\mu}{}_{\hat{3}}&=(-2A_{\hat 3},\, 0,\, 0,\, 1+\Phi')\,.
\end{align}

Taking into account the antisymmetry of the torsion tensor in its first two indices, all of the nonzero components of the structure functions can be obtained in this case from
\begin{equation}\label{K6}
 C_{\hat 0 \hat i}{}^{\hat 0}=-E_{\hat i}+\frac{3}{2}\,\partial_t\, A_{\hat i}\,, \qquad   C_{\hat i \hat j}{}^{\hat 0}=2 \epsilon_{\hat i \hat j \hat k}\, B^{\hat k}\,
\end{equation}
and
\begin{equation}\label{K7}
 C_{\hat i \hat \alpha}{}^{\hat j}=\partial_{\alpha}\, \Phi' \, \delta_{\hat i}^{\hat j}\,.
\end{equation}

To illustrate further the nature of Weitzenb\"ock's torsion, we calculate in this section the structure functions for the natural tetrad frames of the \emph{static} observers in the \emph{exterior} Kerr spacetime. To this end, let us first consider the case of a general stationary metric of the form  
\begin{eqnarray}
\label{K8}
ds^2=g_{00}(dx^0){}^2+g_{11}(dx^1){}^2 +g_{22}(dx^2){}^2+2g_{03}dx^0 dx^3 +g_{33}(dx^3){}^2
\end{eqnarray}
where the metric coefficients $g_{\alpha\beta}$ depend only upon $x^1$ and $x^2$.
A static observer in this case has a 4-velocity vector given by
\beq \label{K9}
e_{\hat 0}=\frac{1}{\sqrt{-g_{00}}}\partial_0\,,
\eeq
and a natural adapted spatial frame that consists of the three vectors
\beq \label{K10}
e_{\hat 1}=\frac{1}{\sqrt{g_{11}}}\partial_1 \,,\quad e_{\hat 2}=\frac{1}{\sqrt{g_{22}}}\partial_2 \,,\quad e_{\hat 3}={\cal F}\,\left(-{\cal G}\,\partial_0 +\partial_3  \right) \,,
\eeq
where
\begin{equation}\label{K11}
{\cal F}:=\frac{1}{\sqrt{g_{33}-\frac{g_{03}^2}{g_{00}}}}\,, \qquad {\cal G}:=\frac{g_{03}}{g_{00}}\,.
\end{equation}
Furthermore, we note that
\begin{equation}\label{K11a}
\sqrt{-g}\,{\cal F}=\sqrt{-g_{00}\,g_{11}\,g_{22}}\,.
\end{equation}

For the structure functions in this case, we have the following general results for the gravitoelectric  components
\begin{equation}\label{K12}
 C_{\hat 0 \hat 1}{}^{\hat 0}=\frac{1}{\sqrt{g_{11}}}\partial_{1} \ln \sqrt{-g_{00}}\,, \quad  C_{\hat 0 \hat 2}{}^{\hat 0}=\frac{1}{\sqrt{g_{22}}}\partial_{2} \ln \sqrt{-g_{00}}\,, \quad C_{\hat 0 \hat 3}{}^{\hat 0}=0\,
\end{equation} 
and the corresponding  gravitomagnetic components
\begin{equation}\label{K13}
C_{\hat i \hat j}{}^{\hat 0}= \sqrt{-g_{00}}\,{\cal F}\, \epsilon_{\hat i \hat j \hat k}\, {\tilde {\cal G}}^{\hat k}\,, 
\qquad {\tilde {\cal G}}^{\hat i}:=  \epsilon^{\hat i \hat j \hat k}\, \partial_{\hat j}\, (0, 0, {\cal G})_{\hat k}
=(\frac{1}{\sqrt{g_{22}}}\,\partial_2{\cal G}, -\frac{1}{\sqrt{g_{11}}}\,\partial_1{\cal G}, 0)\,,
\end{equation}
where $\partial_{\hat i}=e^\mu{}_{\hat i}\, \partial_{\mu}$ is the Pfaffian derivative operator. Moreover, $C_{\hat 0 \hat i}{}^{\hat j}=0$ and all of the other nonzero spatial components can be obtained from 
\begin{equation}\label{K14}
 C_{\hat 1 \hat 2}{}^{\hat 1}=-\frac{1}{\sqrt{g_{22}}}\, \partial_2 \ln \sqrt{g_{11}}\,, \qquad  C_{\hat 2 \hat 1}{}^{\hat 2}=-\frac{1}{\sqrt{g_{11}}}\, \partial_1 \ln \sqrt{g_{22}}\,
\end{equation}
and
\begin{equation}\label{K15}
 C_{\hat 3 \hat 1}{}^{\hat 3}=\frac{1}{\sqrt{g_{11}}}\, \partial_1 \ln {\cal F}\,, \qquad  C_{\hat 3 \hat 2}{}^{\hat 3}=\frac{1}{\sqrt{g_{22}}}\, \partial_2 \ln {\cal F}\,.
\end{equation}

The torsion vector $C_{\alpha}$ can be easily calculated from these results and we find that $C_{\hat \alpha}=(0, C_{\hat 1}, C_{\hat 2}, 0)$, where
\begin{equation}\label{K15a}
 C_{\hat 1} =\frac{1}{\sqrt{g_{11}}}\, \partial_1 \ln {\left(\frac{\sqrt{-g_{00}}}{\sqrt{g_{22}}}\,{\cal F}\right)}\,, \qquad  C_{\hat 2} =\frac{1}{\sqrt{g_{22}}}\, \partial_2 \ln {\left(\frac{\sqrt{-g_{00}}}{\sqrt{g_{11}}}\,{\cal F}\right)}\,.
\end{equation}
Furthermore, the torsion pseudovector $\check{C}_{\alpha}$ is given in this case by
\begin{equation}\label{K15b}
 \check{C}_{\hat \alpha} = (0, \frac{1}{3}\,\sqrt{-g_{00}}\,{\cal F}\,\tilde {\cal G}_{\hat i})\,.
\end{equation}

The metric of the exterior Kerr spacetime is of the general form of Eq.~\eqref{K8} when written in Boyer-Lindquist coordinates $x^\alpha=(t,r,\theta,\varphi)$; in fact, for a Kerr source with mass $M$ and angular momentum $Ma$, we have
\begin{equation}\label{K16}
g_{00}= -1+\frac{2Mr}{\Sigma}\,, \qquad g_{11}=\frac{\Sigma}{\Delta}\,, \qquad g_{22}=\Sigma 
\end{equation}
\begin{equation}\label{K17}
g_{03}=-\frac{2Mra}{\Sigma}\sin^2 \theta\,, \qquad 
g_{33}= \left( r^2+a^2 +\frac{2Mra^2}{\Sigma}\sin^2\theta \right)\sin^2\theta\,.
\end{equation}
Here,
\beq \label{K18}
\Delta=r^2+a^2-2Mr\,,\qquad \Sigma=r^2+a^2\cos^2\theta\,
\eeq
and 
\beq \label{K19}
g_{00}\,g_{33}-g_{03}^2=-\Delta \sin^2 \theta\,.
\eeq

The Kerr structure functions for static observers can be obtained from Eqs.~\eqref{K12}-\eqref{K15} using 
\beq \label{K20}
{\cal F}=\left[\frac{\Sigma-2Mr}{\Sigma \Delta}\right]^{\frac{1}{2}} \frac{1}{\sin \theta}\,, \qquad {\cal G}= \left(\frac{2Mra}{\Sigma-2Mr}\right)\sin^2\theta\,.
\eeq
We can now compare and contrast the frame components of the torsion tensor with those of the corresponding curvature tensor given in Appendix B. 

Further simplifications arise in the Schwarzschild case ($a=0$); that is, the nonzero components of torsion can be obtained from 
\begin{equation}\label{K21}
 C_{\hat 0 \hat 1}{}^{\hat 0}=\frac{M}{r^2 \sqrt{1-\frac{2M}{r}}}\,, \quad  C_{\hat 2 \hat 1}{}^{\hat 2}=C_{\hat 3 \hat 1}{}^{\hat 3}=-\frac{1}{r}\sqrt{1-\frac{2M}{r}}\,,\quad C_{\hat 3 \hat 2}{}^{\hat 3} =- \frac{\cot \theta}{r}\,.
\end{equation}

\section{infinitesimal parallelograms}

Consider two {\it infinitesimal} vectors $A^\mu$ and $B^\mu$ at an event $P$ in spacetime. Suppose that  $A^\mu$ is parallel transported along $B^\mu$ via a general connection $\Gamma$ and $B^\mu$
is in turn parallel transported along $A^\mu$ as in Figure 1. The resulting infinitesimal parallelogram in general suffers from a lack of closure if the connection is not symmetric; in fact, as illustrated in Figure 1, $(CD)^\mu=C_{\alpha \beta}{}^\mu(P) A^\alpha B^\beta$.

\begin{figure}[h]
\begin{center}
\includegraphics[scale=0.6]{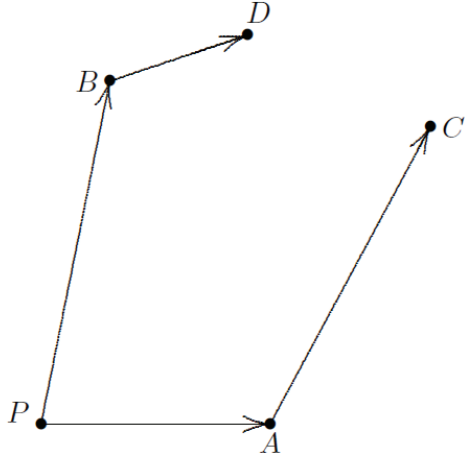}
\end{center}
\caption{
Schematic representation of an infinitesimal parallelogram. Here $(BD)^\mu=A^\mu-\Gamma^\mu_{\alpha \beta} (P)B^\alpha A^\beta$, while $(AC)^\mu=B^\mu-\Gamma^\mu_{\alpha \beta} (P)A^\alpha B^\beta$. Hence $(CD)^\mu=(\Gamma^\mu_{\alpha \beta}-\Gamma^\mu_{\beta \alpha}) A^\alpha B^\beta$.
}
\end{figure}

It is possible to introduce a coordinate system in the neighborhood of event $P$ such that the symmetric part of the connection vanishes~\cite{BaMa}; that is, in the new system of coordinates $\Gamma^\mu_{(\alpha \beta)} (P)=0$. In this case, as depicted in Figure 2, $(CD)^\mu = 2 \delta^\mu$, where $\delta^\mu = \frac{1}{2}C_{\alpha\beta}{}^\mu(P) A^\alpha B^\beta$.

\begin{figure}[h]
\begin{center}
\includegraphics[scale=0.6]{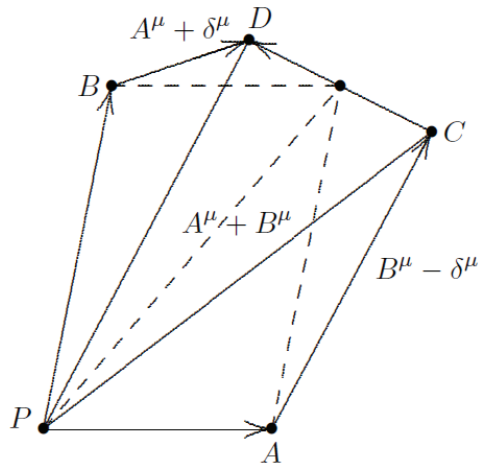}
\end{center}
\caption{
Schematic  representation of the nonclosure of the parallelogram when $\Gamma^\mu_{(\alpha \beta)} (P)=0$..
}
\end{figure}

It follows that in our extended GR framework, if the Weitzenb\"ock torsion does not vanish at an event $P$, then infinitesimal parallelograms based at $P$ do not close. The situation is different, however, for non-infinitesimal parallelograms, whose closure, or lack thereof, would crucially depend on the detailed circumstances at hand and the nature of the spacetime under consideration.

\section{Concluding Remarks}

Torsion, like curvature, is a basic tensor associated with a linear connection. The measurement of spacetime torsion depends upon the role that the torsion field plays in the physical theory. In the context of the Poincar\'e gauge theory of gravitation, for instance, Cartan's torsion is related to intrinsic spin and the possibility of its measurement has been explored in that framework~\cite{He, La, HOP}. Another approach involves the motion of extended bodies in the context of nonminimal theories, where torsion couplings can be important~\cite{PO}. 

Weitzenb\"ock's torsion has been previously studied in the context of teleparallelism~\cite{MVRN, MUF}. In this paper, we have generalized previous work on the physical aspects of Weitzenb\"ock's torsion~\cite{BaMa}. In our extended GR framework, we have studied the general properties of this torsion field for orthonormal frames that are naturally adapted to static observers in gravitational fields. For the measured components of the torsion tensor, $C_{\hat \alpha \hat \beta}{}^{\hat \gamma}$, we find that $C_{\hat 0 \hat i}{}^{\hat 0}$ represents what is essentially the gravitoelectric field, while $C_{\hat i \hat j}{}^{\hat 0}$ represents what is essentially the gravitomagnetic field. Moreover, $C_{\hat 0 \hat i}{}^{\hat j}$ is related to the 
nonstationary character of the gravitational field and $C_{\hat i \hat j}{}^{\hat k}$ has in general mixed properties involving both the gravitoelectric and gravitomagnetic aspects.  These results should be compared and contrasted with the frame components of the curvature tensor.  Our work illustrates the fact that in the extended GR framework, curvature and torsion are complementary representations of the gravitational field.

\begin{acknowledgments}
We are grateful to Friedrich Hehl and Jos\'e Maluf for valuable discussions. 
\end{acknowledgments}

\appendix %{}

\section{Nonsymmetric Connections}\label{appA}

In our convention, the covariant derivative associated with a general nonsymmetric connection $\Gamma^\mu_{\alpha \beta}$ is defined for vector fields $A^\mu$ and $B_\mu$ as
\begin{equation}\label{A1}
\nabla_\alpha \, A^\mu= \partial_\alpha\, A^\mu + \Gamma^\mu_{\alpha \beta}\, A^\beta\,, \qquad \nabla_\alpha \, B_\mu= \partial_\alpha\, B_\mu - \Gamma^\beta_{\alpha \mu}\, B_\beta\,.
\end{equation}
It follows that $\nabla_\nu\, e_{\mu}{}^{\hat \alpha}=0$ for the  Weitzenb\"ock connection. Furthermore, for a covariant vector field $A_\mu$, 
\begin{equation}\label{A1a}
\nabla_\mu\, A_\nu-\nabla_\nu\, A_\mu= \partial_\mu\, A_\nu - \partial_\nu\, A_\mu -C_{\mu \nu}{}^\alpha\, A_\alpha\,.
\end{equation}

For a scalar field $S$, $\nabla_\alpha\,S=\partial_\alpha\,S$ and we have
\begin{equation}\label{A2}
(\nabla_\alpha\, \nabla_\beta-\nabla_\beta\, \nabla_\alpha)\, S=C_{\alpha \beta}{}^\mu\, \partial_\mu\, S\,;\end{equation}
moreover,  the \emph{Ricci identity} takes the form
\begin{equation}\label{A3}
(\nabla_\alpha\, \nabla_\beta-\nabla_\beta\, \nabla_\alpha)\, A_\mu= R^\gamma{}_{\mu \alpha \beta} A_\gamma + C_{\alpha \beta}{}^\nu\, \nabla_\nu\,A_\mu\,.
\end{equation}
Here,
\begin{equation}\label{A4}
 R^\gamma{}_{\mu \alpha \beta}= -  R^\gamma{}_{\mu \beta \alpha}
 \end{equation}
 is the curvature tensor given by
 \begin{equation}\label{A5}
 R^\alpha{}_{\mu \beta \nu}= \partial_\beta\, \Gamma^\alpha_{\nu \mu}-\partial_\nu\, \Gamma^\alpha_{\beta \mu}+\Gamma^\alpha_{\beta \sigma}\, \Gamma^\sigma_{\nu \mu}-\Gamma^\alpha_{\nu \sigma}\, \Gamma^\sigma_{\beta \mu}\,,
 \end{equation}
 which vanishes in the case of Weitzenb\"ock's connection.

\section{Frame components of the Riemann tensor for static observers in the exterior Kerr spacetime}\label{appB}

The symmetries of the Riemann curvature tensor make it possible to express its frame components $^0R_{\hat \alpha \hat \beta \hat \gamma \hat \delta}$ as elements of a symmetric $6\times 6$ matrix $^0{\mathcal R}= (^0{\mathcal R}_{IJ})$, where $I$ and $J$ range over the set $(01, 02, 03, 23, 31, 12)$; that is, 
\beq \label {B1}
^0{\mathcal R}=\left[  
\begin{array}{cc}
{\hat {\mathcal E}} & {\hat {\mathcal B}}\cr
{\hat {\mathcal B}}^T & {\hat {\mathcal S}}\cr
\end{array}
\right]\,,
\eeq
where ${\hat {\mathcal E}}$ and ${\hat {\mathcal S}}$ are symmetric $3\times 3$ matrices and ${\hat {\mathcal B}}$ is traceless. Here  ${\hat {\mathcal E}}$ and  ${\hat {\mathcal B}}$ correspond to the gravitoelectric and gravitomagnetic components of spacetime curvature, respectively, while  ${\hat {\mathcal S}}$ corresponds to its  spatial components. In a Ricci-flat spacetime,  ${\hat {\mathcal S}}=- {\hat {\mathcal E}}$,  ${\hat {\mathcal E}}$ is traceless and  ${\hat {\mathcal B}}$ is symmetric.   

For the family of static observers in the exterior Kerr spacetime, the nonvanishing components of the symmetric and traceless  ${\hat {\mathcal E}}=({\mathcal E}_{\hat i \hat j})$  and  ${\hat {\mathcal B}}=({\mathcal B}_{\hat i \hat j})$ with respect to the frame~\eqref{K9}-\eqref{K10} can be obtained from  
\begin{equation} \label{B2}
{\mathcal E}_{\hat 1 \hat 1}=-\frac{Mr (r^2-3a^2\cos^2\theta)(2\Delta + a^2\sin^2 \theta)}{\Sigma^3 (\Sigma - 2Mr)}\,, \quad  
{\mathcal E}_{\hat 1 \hat 2}=-\frac{3Ma^2(3r^2-a^2\cos^2\theta)\sqrt{\Delta}\cos \theta \sin \theta}{\Sigma^3 (\Sigma -2Mr)}\,,
\end{equation}
\begin{equation} \label{B3}
{\mathcal E}_{\hat 2 \hat 2}= \frac{Mr (r^2-3a^2\cos^2\theta)(\Delta+2a^2\sin^2 \theta) }{\Sigma^3 (\Sigma -2Mr)}\,, \quad 
{\mathcal E}_{\hat 3 \hat 3}= \frac{Mr}{\Sigma^3}(r^2-3a^2\cos^2\theta)\,
\end{equation}
and
\beq \label{B4}
{\mathcal B}_{\hat 1 \hat 1}=\rho \, {\mathcal E}_{\hat 1 \hat 1}\,, \quad    
{\mathcal B}_{\hat 1 \hat 2}=\frac{1}{\rho}\, {\mathcal E}_{\hat 1 \hat 2}\,, \quad
{\mathcal B}_{\hat 2 \hat 2}= \rho\,{\mathcal E}_{\hat 2 \hat 2}\,, \quad
{\mathcal B}_{\hat 3 \hat 3}=\rho \, {\mathcal E}_{\hat 3 \hat 3}\,,
\eeq
where the dimensionless ratio $\rho$ is given by
\beq \label{B5}
\rho = \frac{a\cos\theta}{r} \left(\frac{3r^2-a^2\cos^2\theta}{r^2-3a^2\cos^2\theta}\right)\,.
\eeq
    We note that ${\hat {\mathcal E}}$ and ${\hat {\mathcal B}}$ diverge at the stationary limit $(g_{00}=0)$, and ${\hat {\mathcal B}}=0$ in the nonrotating Schwarzschild case, where $a=0$. 

\section{Weitzenb\"ock torsion invariants}

Let $e_{\hat \alpha}$ be a given orthonormal frame and $C_{ \hat \alpha \hat \beta }{}^{\hat \gamma}$ be the associated frame components of the structure functions.
It is convenient to introduce the notation
\beq \label{C1}
C_{\hat 0 \hat a}{}^{\hat 0}=\Psi_{\hat a}\,,\qquad C_{\hat a \hat b}{}^{\hat 0}=\epsilon_{\hat a \hat b \hat c}W^{\hat c}\,,\qquad
C_{\hat 0 \hat a}{}^{\hat b}=\chi_{\hat a}{}^{\hat b}\,.
\eeq
Then, the {\it torsion vector}  $C_{\hat \alpha}$ has components
\beq \label{C2}
C_{\hat 0}=-\chi_{\hat i}{}^{\hat i}\,,\qquad C_{\hat a}=\Psi_{\hat a}-C_{ \hat a \hat b }{}^{\hat b}\,;
\eeq
while the  {\it torsion pseudovector} $\check{C}_{\hat \alpha}$ has components
\beq \label{C3}
\check{C}_{\hat 0}=-\frac16\, \epsilon^{\hat a \hat b}{}_{\hat c}\, C_{\hat a \hat b}{}^{\hat c}\,,\qquad 
\check{C}_{\hat a}=\frac13 \,(W_{\hat a} -\epsilon_{\hat a \hat b \hat c}\, \chi^{\hat b \hat c})\,.
\eeq

Next, we consider the three algebraic Weitzenb\"ock  invariants of the torsion tensor, namely,
\beq \label{C4}
{\mathcal I}_1=C_{\hat \alpha \hat \beta}{}^{\hat \gamma} C^{\hat \alpha \hat \beta}{}_{\hat \gamma}\,,\qquad
{\mathcal I}_2=C_{\hat \alpha \hat \beta \hat \gamma} C^{ \hat \gamma\hat \beta\hat \alpha}\,,\qquad
{\mathcal I}_3=C_{\hat \alpha}C^{\hat \alpha}\,.
\eeq
In terms of the components of the torsion tensor, we have for ${\mathcal I}_1$ and ${\mathcal I}_2$
\begin{eqnarray} \label{C5}
{\mathcal I}_1&=& 2 \Psi_{\hat a}\Psi^{\hat a}-2 W_{\hat a}W^{\hat a} -2\chi_{\hat a \hat b}\, \chi^{\hat a\hat b}+C_{\hat a \hat b}{}^{\hat c}C^{\hat a \hat b}{}_{\hat c}\,,\nonumber\\
{\mathcal I}_2&=& \Psi_{\hat a}\Psi^{\hat a}-2 \epsilon_{\hat a \hat b \hat c}\,W^{\hat a} \chi^{\hat b \hat c}-\chi_{\hat a \hat b}\, \chi^{\hat b \hat a}+C_{\hat a \hat b}{}^{\hat c}C_{\hat c}{}^{\hat b \hat a}\,.
\end{eqnarray}
Simplifications occur either in the case of the static  Fermi observers, at the order of approximation employed in section II, or in the case of static observers in the general stationary axisymmetric spacetime considered in section III, since $\chi_{\hat a \hat b}=0$ in these cases.

\end{document}